# MULTIPLICATION METHOD FOR FACTORING NATURAL NUMBERS

# ФАКТОРИЗАЦИЯ НАТУРАЛЬНЫХ ЧИСЕЛ МЕТОДОМ УМНОЖЕНИЯ

**Igor Nesiolovskiy**
*Podolsk, Moscow Region, Russia*
nesiolovskiy@gmail.com

**Artem Nesiolovskiy**
*Irvine, CA, USA*
yinmute@gmail.com

**Игорь Несиоловский**
*Россия, Московская область, Подольск*
nesio@mail.ru

**Артем Несиоловский**
*США, Калифорния, Ирвайн*
yinmute@gmail.com

**Abstract (Аннотация)**

We offer multiplication method for factoring big natural numbers which extends the group of the Fermat's and Lehman's factorization algorithms and has run-time complexity $O(n^{1/3})$. This paper is argued the finiteness of proposed algorithm depending on the value of the factorizable number $n$. We provide here comparative tests results of related algorithms on a large amount of computational checks. We describe identified advantages of the proposed algorithm over others. The possibilities of algorithm optimization for reducing the complexity of factorization are also shown here.

Мы предлагаем алгоритм факторизации больших натуральных чисел методом умножения, который расширяет группу алгоритмов факторизации Ферма и Лемана и характеризуется сложностью выполнения $O(n^{1/3})$. В данной статье обосновывается конечность его работы в зависимости от величины факторизуемого числа $n$. Представлены результаты сравнительных испытаний родственных алгоритмов на большом объеме вычислительных тестов и продемонстрированы выявленные преимущества предлагаемого алгоритма перед другими. Показаны возможности оптимизации алгоритма для снижения трудоемкости факторизации.

## 1 Введение (Introduction)

В программных приложениях теории чисел и защиты информации актуальна задача быстрой факторизации больших целых чисел. В данной работе предлагается наш подход к ее решению методом умножения со сложностью алгоритма факторизации $O(n^{1/3})$. Метод умножения (сокращенно MMFFNN – multiplication method for factoring natural numbers) является родственным с методами факторизации Ферма и Лемана и расширяет их семейство, в которых у Ферма циклично используется сложение, а у Лемана комбинируется и циклично применяется умножение и сложение. В MMFFNN мы будем циклично использовать умножение.

Вероятно, еще Ферма знал, что предложенный им метод факторизации, основанный на представлении числа в виде разности двух квадратов, можно оптимизировать по скорости выполнения для случая, когда простые числа $a$ и $b$, произведение которых равно $n$, достаточно удалены от значения $\sqrt{n}$. Идея подобной оптимизации состоит в том, чтобы подбирать такой множитель $u * v$ к числу $n$, в котором отношение $u/v$ близко к отношению $a/b$, и использовать его в алгоритме Ферма. Первый способ такого подбора был предложен в работе [1]. Ниже мы рассмотрим два других способа: *простого умножения* (Simple Multiplication – SM), близкий к которому описывался в работе [2], и *рекурсивного умножения* (Recursive Multiplication – RM). А также сравним все три способа между собой по результатам проведенных с ними вычислительных экспериментов.

## 2 Обоснование алгоритма (Rationale behind the algorithm)

В данном разделе мы выполним обоснование алгоритма RM (Recursive Multiplication) метода MMFFNN.

Положим $N = A * B = m * n * k$, где $n = a * b$, $a$ и $b$ – нечетные простые и такие, что

$$n^{1/3} < a \leq b < n^{2/3},$$

$m$ – небольшой по сравнению с $n$ целочисленный множитель, $k = k_1 * k_2 * \dots * k_g$, где $g = \lfloor \log_3(\log_2(m * n)) \rfloor$ (с округлением результата логарифмирования по основанию 3 вниз до целого), $k_i$ – натуральные и такие, что $1 < k_i < \left\lfloor (m * n)^{(1/3)^i} \right\rfloor$ (с округлением степенного выражения вниз до целого). Допустим пока, что $m = 1$. Примем $D = B - A$, где $D \geq 0$, и выразим $A$ в виде $A = \frac{\sqrt{D^2 + 4 * N} - D}{2}$. Освободим $D$ из под квадратного корня:

$$D = \sqrt{\left\lceil \sqrt{4 * n * k} \right\rceil^2 - 4 * n * k}$$

(с округлением $\sqrt{4*n*k}$ вверх до целого). Тогда $A = \frac{\lceil\sqrt{4*n*k}\rceil - D}{2}$. Теперь если при некотором искомом $k$ значение $D$ оказывается целым, то $a = gcd(n, A)$ или $b = gcd(n, A)$, т.е. факторизация $n$ выполнена при условии, что $gcd(n, A) \neq 1$ и $gcd(n, A) \neq n$ (здесь $gcd$ – greatest common divider).

Очевидно, что чем меньше значение $k$, тем выше эффективность алгоритма факторизации. Оценим границу искомой величины $k$. Из выражения для $D$ запишем

$$\lceil\sqrt{4*n*k}\rceil^2 - D^2 = 4*n*k = \left(\sqrt{4*a*b*k_a*k_b}\right)^2 =$$
$$(a*k_a + b*k_b)^2 - (a*k_a - b*k_b)^2,$$

где $A = a*k_a$, $B = b*k_b$. При этом справедливо, что

$$2*\sqrt{4*a*b*k_a*k_b} \le (a*k_a - b*k_b)^2 = (a*k_a + b*k_b)^2 - \left(\sqrt{4*a*b*k_a*k_b}\right)^2.$$

В теории чисел известен результат (рассмотренный, например, в [3]): для простых нечетных $p$ и $q$ при условиях

$$(p*q)^{1/3} < p \le q < (p*q)^{2/3}$$

существуют такие значения целых $r$ и $s$, что

$$r*s < (p*q)^{1/3} \quad \text{и} \quad |p*r - q*s| < (p*q)^{1/3}.$$

Используем его для наших условий.

Если $g = 1$, то справедливо, что

$$|a*k_{1a} - b*k_{1b}| < n^{1/3}, \quad k_1 = k_{1a}*k_{1b} < n^{1/3}.$$

С учетом приведенного выше неравенства для выражения $(a*k_a - b*k_b)^2$ допустимо записать следующее:

$$2*\sqrt{4*n*k_{1a}*k_{1b}} \le (a*k_{1a} - b*k_{1b})^2 < (n^{1/3})^2.$$

Откуда верно, что $k_1 < n^{1/3}$.

Если $g = 2$, то справедливо, что

$$|k_{1a}*k_{2a} - k_{1b}*k_{2b}| < n^{1/9}, \quad k_2 = k_{2a}*k_{2b} < n^{1/9}.$$

Аналогично получаем:

$$2*\sqrt{4*\ n^{1/3}*k_{2a}*k_{2b}} \le (k_{1a}*k_{2a} - k_{1b}*k_{2b})^2 < (n^{1/9})^2.$$

Откуда верно, что $k_2 < n^{1/9}$.

Если $g = 3$, также справедливо, что

$$|k_{2a}*k_{3a} - k_{2b}*k_{3b}| < n^{1/27}, \quad k_3 = k_{3a}*k_{3b} < n^{1/27}$$

Аналогично получаем:

$$2 * \sqrt{4 * n^{1/9} * k_{3a} * k_{3b}} \leq (k_{2a} * k_{3a} - k_{2b} * k_{3b})^2 < (n^{1/27})^2 .$$

Откуда верно, что $k_3 < n^{1/27}$ .

Поскольку при конечном $n$ индекс $i$ имеет конечное значение, допустим, что заданное для факторизации число $n$ таково, что $g = 3$. В этом случае справедливо, что

$$k = k_1 * k_2 * k_3 < n^{1/3} * n^{1/9} * n^{1/27}$$

при факторизации чисел $n < 2^{27} = 134217728$, т.к. $1 < k_3 = n^{1/27}$.
Если $n$ таково, что $g = 4$, то в этом случае справедливо, что

$$k = k_1 * k_2 * k_3 * k_4 < n^{1/3} * n^{1/9} * n^{1/27} * n^{1/81}$$

при факторизации чисел $n < 2^{81} = 2417851639229258349412352$, т.к. $1 < k_4 = n^{1/81}$.
В случае, если $n \to \infty$. то $k \to n^{1/2}$ – в силу свойств геометрической прогрессии с элементами $i$.

Таким образом в алгоритме RM метода MMFFNN величина $k$ для заданного $n$ конечна. Если при достижении предельного значения $k$ алгоритм не факторизует число $n$, то оно является простым. Отметим, что выше обоснована вторая фаза алгоритма, которая выполняется в случае, если первая фаза безрезультатна – перебора возможных простых делителей $c \leq n^{1/3}$ числа $n$. Поэтому двухфазный алгоритм факторизации RM имеет сложность выполнения $O\ (n^{1/3})$.

## 3 Реализация и испытания алгоритма (Implementation and testing of the algorithm)

Для проведения вычислительных экспериментов с методом MMFFNN были реализованы два основных программных приложения на платформе Java: *генерации* чисел требуемого вида и их *факторизации*.

Приложение генерации формирует наборы целых составных чисел $n = a * b$ заданной в десятичном виде разрядности $r$ для $n$, где $a$ случайное вероятно простое число в пределах $n^{1/3} < a < n^{1/2}$, $b$ случайное вероятно простое число, согласованное с $r$.

Приложение факторизации находит факторы разложения чисел на множители $a$ и $b$ из предварительно сгенерированных наборов составных чисел $n$ по трем алгоритмам: Лемана, SM (Simple Multiplication) и RM (Recursive Multiplication).

Реализованный в приложении факторизации алгоритм Лемана работает согласно следующему машинному псевдокоду (см. обоснование алгоритма в [3]):

<u>Lehman's algorithm</u>

**input** $n$

$iteration \leftarrow 0; factor \leftarrow 0$

**for** $k \leftarrow 1$ ***to*** $\lfloor n^{1/3} \rfloor$ **do**

  **for** $d \leftarrow 1$ ***to*** $\lfloor n^{1/6}/(4 * k^{1/2}) \rfloor + 1$ **do**

    $iteration \leftarrow iteration + 1$

    **if** $((\lceil \sqrt{4 * n * k} \rceil + d)^2 - 4 * n * k)^{1/2} = integer$ **then**

      $factor \leftarrow gcd\left(n, \left(\lceil \sqrt{4 * n * k} \rceil + d\right) - ((\lceil \sqrt{4 * n * k} \rceil + d)^2 - 4 * n * k)^{1/2}\right)$

      **if** $1 < factor < n$ **then** $return\ (f, iteration)$ **else**

        $factor \leftarrow gcd\left(n, \left(\lceil \sqrt{4 * n * k} \rceil + d\right) + ((\lceil \sqrt{4 * n * k} \rceil + d)^2 - 4 * n * k)^{1/2}\right)$

        **if** $1 < factor < n$ **then** $return\ (iteration, factor)$

Реализованный в приложении факторизации алгоритм SM работает согласно следующему машинному псевдокоду (см. обоснование алгоритма в [2]).

<u>Simple Multiplication algorithm</u>

**input** $n, m$

$iteration \leftarrow 0; factor \leftarrow 0$

**for** $k \leftarrow 1$ ***to*** $infinity$ **do**

  $iteration \leftarrow iteration + 1$

  **if** $(\lceil \sqrt{m * n * k} \rceil^2 - m * n * k)^{1/2} = integer$ **then**

    $factor \leftarrow gcd\left(n, \lceil \sqrt{m * n * k} \rceil - (\lceil \sqrt{m * n * k} \rceil^2 - m * n * k)^{1/2}\right)$

    **if** $1 < factor < n$ **then** $return\ (iteration, factor)$

Реализованный в приложении алгоритм RM работает согласно следующему машинному псевдокоду.

<u>Recursive Multiplication algorithm</u>

**input** $n, m$ $'factor -\ sought\text{-}for\ factor\ of\ n$

$iteration \leftarrow 0; factor \leftarrow 0;$ $'i - recursion\ level,\ k - loop\ variable\ of\ level$

$i_{out} \leftarrow 0;\quad k_{out} \leftarrow 0;\quad k_mult_{out} \leftarrow 1$ $'k_{mult} - product\ of\ loop\ variables$

**call** $recursion\ (n, m, iteration, i_{out}, k_{out}, k_mult_{out}, factor)$

$return\ (iteration, factor)$

**procedure** $recursion\ (n, m, iteration, i_{in}, k_{in}, k_mult_{in}, factor)$

**if** $i_{in} = 0$ **then** $i_{out} \leftarrow \lfloor \log_3(\log_2(m * n)) \rfloor$

**if** $i_{in} > 1$ **then** $i_{out} \leftarrow i_{in} - 1$

**if** $i_{in} = 1$ **then** $i_{out} \leftarrow i_{in}$

$k_lim_{out} \leftarrow \left\lfloor (m * n)^{(1/3)^{i_{out}}} \right\rfloor$

**for** $k_{out} \leftarrow k_{in}$ ***to*** $k_lim_{out}$ **do**

$k_mult_{out} \leftarrow k_mult_{in} * k_{out}$

**if** $i_{in} > 1$ **then**

**if** $isSieved(i_{in}\ ,\ k_{in}\ ,\ i_{out}\ ,\ k_{out}\ ,\ k_lim_{out}) = true$ **then**

**call** $recursion\ (n, m, iteration, i_{out}, k_{out}, k_mult_{out}, factor)$

**else**

$iteration \leftarrow iteration + 1$

**if** $(\lceil\sqrt{4 * m * n * k_mult_{out}}\rceil^2 - 4 * m * n * k_mult_{out})^{1/2} = integer$ **then**

$$factor \leftarrow gcd\left(n, \lceil\sqrt{4 * m * n * k_mult_{out}}\rceil - (\lceil\sqrt{4 * m * n * k_mult_{out}}\rceil^2 - 4 * m * n * k_mult_{out})^{1/2}\right)$$

**if** $1 < factor < n$ **then** $return\ (iteration, factor)$

Для оценки трудоемкости исполнения алгоритмов факторизации в каждом из них вычисляется инкрементируемая переменная $iteration$, которая фактически равна сумме проверок ключевого промежуточного числа, итерационно вычисляемого в алгоритме при факторизации числа $n$, на соответствие полному квадрату.

Кроме того в алгоритме RM использована привилегия, присущая именно этому методу – функция $isSieved$, с помощью которой реализуется специальное решето, отсеивающее избыточные итерации факторизации, в которых дублируется значение рекурсивного множителя $k = k_1 * k_2 * ... * k_s$ (например, при k=24=1*2*12=1*3*8=2*2*6=2*3*4, если s=3), что замедляет рост значения переменной $iteration$. Без подобного решета, т.е. без строки **if** $isSieved(i_{in}\ ,\ k_{in}\ ,\ i_{out}\ ,\ k_{out}\ ,\ k_lim_{out}) = true$ **then** в псевдокоде, алгоритм будет работать, но несколько менее эффективно.

## 4 Анализ результатов испытаний алгоритма (Analysis of the test results of the algorithm)

Исследование алгоритма RM (Recursive Multiplication) метода MMFFNN проводилось с целью анализа эффективности его выполнения на компьютере в сравнении с другими родственными алгоритмами – Лемана и SM (Simple Multiplication).

Для исследования были сгенерированы 14 наборов составных чисел $n$ разрядностью $r$ от 3 до 16 с шагом 1, задаваемой в десятичном виде, по 20000 чисел в каждом и с указанными в предыдущем разделе ограничениями на сомножители $a$ и $b$. При генерации чисел контроль их дублирования в наборе не предусматривался.

При выполнении факторизации чисел по каждому из трех алгоритмов осуществлялся дополнительный контроль соответствия найденных факторов разложения множителям $a$ и $b$, использованным при генерации чисел $n$.

В алгоритме SM согласно рекомендаций источника [2] было применено значение постоянного множителя $m = 480$. Для полной сопоставимости результатов испытаний в алгоритме RM было применено значение множителя $m = 120$, которое при умножении на константу 4, предусмотренную в алгоритме RM, дает значение совокупного постоянного множителя $4 * m = 480$.

Оценка трудоемкости работы каждого из трех алгоритмов определялась по каждому сгенерированному набору чисел по следующему среднеарифметическому показателю:

$$\overline{q}_r = \left\lfloor \left(\sum_{1}^{20000} iteration_j\right)/20000 \right\rfloor,$$

где $iteration_j$ – суммарное количество попыток алгоритма по извлечению квадратного корня с целью тестирования вычисляемого радикала на получение целочисленного значения при разложении очередного $j$-го числа из набора чисел разрядности $r$, сгенерированного для факторизации.

Результаты испытаний трех алгоритмов представлены ниже в форме таблицы и диаграммы.

Таблица 1. Сравнение результатов испытаний трех алгоритмов

<table>
<tr><th rowspan="2">$r$</th><th colspan="3">$\overline{q}_r$</th></tr>
<tr><th>Lehman</th><th>Simple Multiplication</th><th>Recursive Multiplication</th></tr>
<tr><td>3</td><td>3</td><td colspan="2">1</td></tr>
<tr><td>4</td><td>6</td><td colspan="2">3</td></tr>
<tr><td>5</td><td>14</td><td colspan="2">6</td></tr>
<tr><td>6</td><td>29</td><td colspan="2">13</td></tr>
<tr><td>7</td><td>62</td><td colspan="2">25</td></tr>
<tr><td>8</td><td>135</td><td colspan="2">52</td></tr>
<tr><td>9</td><td>294</td><td colspan="2">110</td></tr>
<tr><td>10</td><td>634</td><td colspan="2">241</td></tr>
<tr><td>11</td><td>1384</td><td colspan="2">517</td></tr>
<tr><td>12</td><td>2955</td><td>1136</td><td>1134</td></tr>
<tr><td>13</td><td>6250</td><td>2503</td><td>2482</td></tr>
<tr><td>14</td><td>13647</td><td>5360</td><td>5323</td></tr>
<tr><td>15</td><td>29490</td><td>11257</td><td>11200</td></tr>
<tr><td>16</td><td>63813</td><td>24604</td><td>24450</td></tr>
</table>

Диаграмма 1. Сравнение результатов испытаний трех алгоритмов

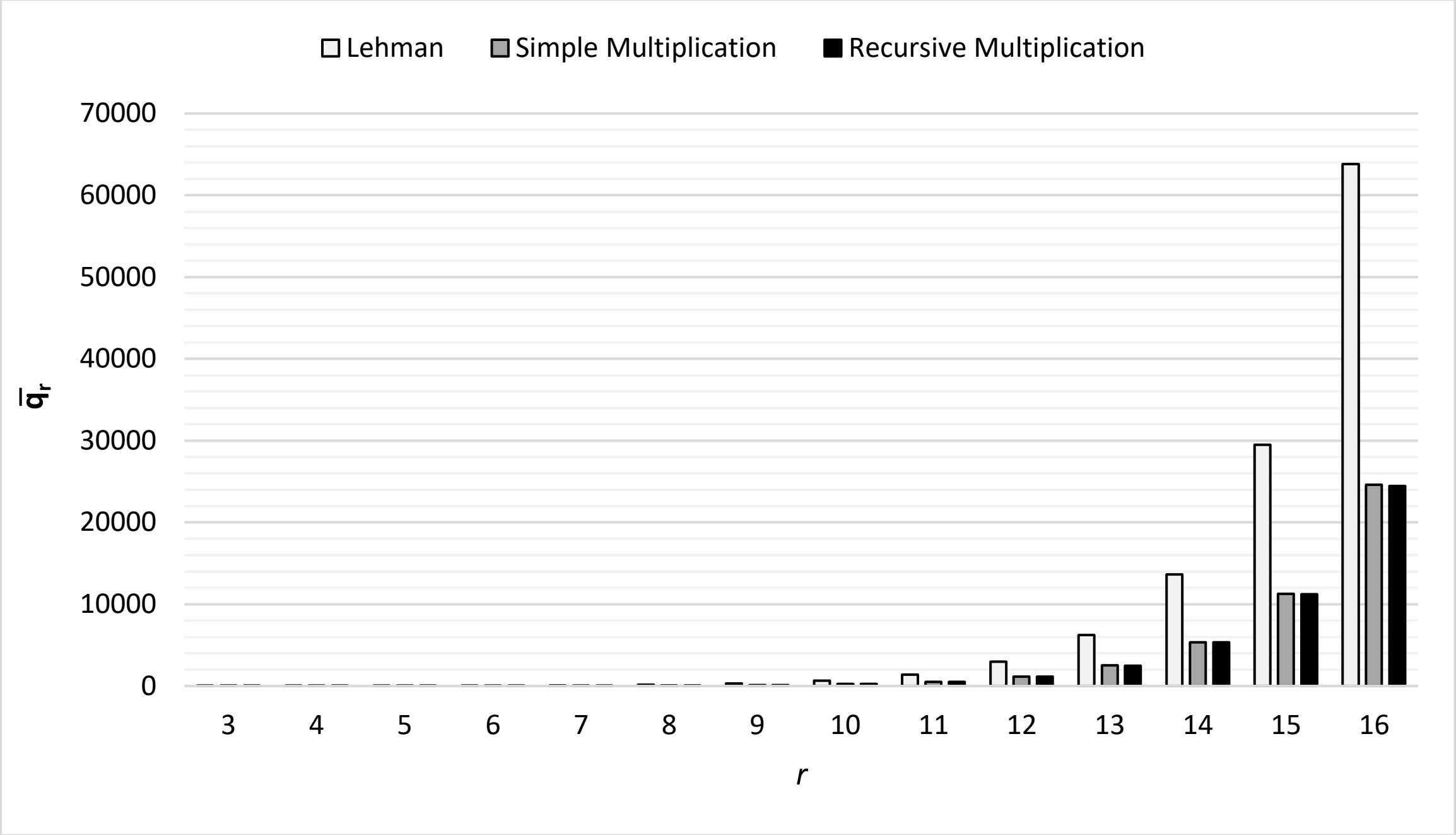

Результаты экспериментов свидетельствуют о следующем.

Все алгоритмы обеспечили корректное выполнение факторизации всех чисел из всех 14-ти тестовых наборов. При этом алгоритмы Лемана и RM уложились в установленные для них теоретические ограничения итерационных циклов поиска значения множителя $k$, алгоритм SM выполнил факторизацию при отсутствии верхней границы цикла подбора значения множителя $k$.

Все алгоритмы показали рост трудоемкости $\overline{q}_r$ факторизации чисел $n$ с увеличением их разрядности $r$ в виде, близком к геометрической прогрессии: алгоритмы SM и RM со знаменателем прогрессии, примерно равным 2, Лемана – со знаменателем большим, чем 2.

Алгоритмы группы MMFFNN (т.е. SM и RM) продемонстрировали в среднем по всем тестовым наборам данных экономию трудоемкости факторизации, основанную на показателе $\overline{q}_r$, в размере более чем 250% в сравнении с аналогичной средней трудоемкостью алгоритма Лемана.

На тестовых наборах факторизуемых чисел разрядности $r$ в диапазоне $3 \le r \le 11$ алгоритмы группы MMFFNN проявили одинаковые результаты трудоемкости разложения чисел, но с дальнейшим ростом $r$ трудоемкость алгоритма SM начала постепенно превышать трудоемкость RM. Этот же эффект будет наблюдаться и для других значений множителей вида $m_{SM} = 4 * m_{RM}$ в алгоритмах SM и RM. Здесь можно предложить следующую оценку десятичной разрядности $r_*$ числа $n$, начиная с которой трудоемкость $\overline{q}_r$ факторизации алгоритма SM будет гарантированно превышать трудоемкость RM с множителями $m_{SM} = 4 * m_{RM}$:

$$r_* = \lceil 6 * \log_{10} m_{RM} \rceil.$$

Она вытекает из следствия приведенного выше обоснования алгоритма RM, относящегося к алгоритму SM: искомый множитель $k$ в алгоритме SM имеет верхнюю границу $k < n^{1/2} = n^{1/3} * n^{1/6}$.

Применяемое значение множителя $m$ в алгоритме RM влияет на трудоемкость $\overline{q}_r$ факторизации, о чем свидетельствует следующий пример сравнения результатов факторизации чисел из описанных выше 14-ти тестовых наборов с разными значениями $m$:

Таблица 2. Сравнение результатов испытаний алгоритма RM с разными значениями $\boldsymbol{m}$

| $r$ | $\overline{q}_r$ | |
|---|---|---|
| | Recursive Multiplication | |
| | $m = 120 = 5!$ | $m = 5040 = 7!$ |
| 3 | 1 | 1 |
| 4 | 3 | 3 |
| 5 | 6 | 6 |
| 6 | 13 | 12 |
| 7 | 25 | 25 |
| 8 | 52 | 52 |
| 9 | 110 | 109 |
| 10 | 241 | 225 |
| 11 | 517 | 461 |
| 12 | 1134 | 979 |
| 13 | 2482 | 2093 |
| 14 | 5323 | 4349 |
| 15 | 11200 | 9461 |
| 16 | 24450 | 20316 |

Применение множителя $m > 1$ в алгоритме RM эффективно не только для интегрального снижения трудоемкости $\overline{q}_r$ факторизации множества чисел, как видно из примера в Таблице 2, но и в частных случаях разложения сложных для алгоритма RM чисел при $m = 1$, например:

$$9\,441\,101\,419\,801 = 2\,174\,023 * 4\,342\,687,$$

$$96\,864\,103\,649\,179 = 5\,680\,679 * 17\,051\,501,$$

$$99\,968\,287\,751\,261 = 9\,994\,573 * 10\,002\,257.$$

Пример из Таблицы 2 демонстрирует непропорциональную зависимость показателя $\overline{q}_r$ от множителя $m$ с ростом значения $r$. При выборе оптимального значения $m$ в алгоритме SM следует выбирать компромисс между негативными факторами, т.е. дополнительными вычислительными затратами на умножение при растущем $m$ и увеличением

$$g = \lfloor \log_3(\log_2(m * n)) \rfloor \quad \text{и} \quad k = k_1 * k_2 * \ldots * k_g,$$

и благоприятными факторами – увеличением количества простых делителей $m$.

## 5 Заключение (Conclusion)

Мы рассмотрели предлагаемый алгоритм факторизации целых чисел RM (Simple Multiplication) метода MMFFNN в сравнении с родственными алгоритмами Лемана и SM (Simple Multiplication). Обосновали конечность его работы в зависимости от величины факторизуемого числа $n$ и сложность его исполнения $O(n^{1/3})$. Проанализировали результаты сравнительных испытаний алгоритмов RM, SM и Лемана на большом объеме вычислительных тестов и продемонстрировали выявленные преимущества алгоритма RM перед другими. Показали возможности оптимизации алгоритма RM для снижения трудоемкости факторизации. Также представили некоторые свойства алгоритма SM, не описанные в [2].

Заметим, что дополнительные приёмы ускорения факторизации, предлагаемые в [1] для алгоритма Лемана и в [2] для алгоритма SM, применимы и для алгоритма RM.

Мы допускаем, что для одних алгоритм RM может представлять некоторую теоретическую значимость, другие могут воспользоваться им в образовательных целях, а третьи попробуют применить его для решения практических задач, возможно в комплексе с другими методами факторизации больших целых чисел.

## 6 Ссылки (References)